\documentclass[aps,prb,twocolumn,reprint,superscriptaddress,citeautoscript]{revtex4-2}
\usepackage[dvipdfmx]{graphicx}
\usepackage{subfigmat}
\usepackage[whole]{bxcjkjatype}

\usepackage{amsmath,amssymb}
\usepackage{color}
\usepackage{bm}
\usepackage{physics}
\usepackage{hyperref}
\hypersetup{
	citecolor=blue,
	colorlinks = true,
	urlcolor = blue
}

\usepackage{comment}
\usepackage[normalem]{ulem}

\usepackage{upgreek}

\usepackage{xcolor}
\definecolor{honey}{HTML}{ec9706}

\begin{document}

\title{Quantum fluctuation in rotation velocity of a levitated magnetic particle}

\author{T. Sato}
\affiliation{Institute for Solid State Physics, The University of Tokyo, Kashiwa, Japan}

\author{Daigo Oue}
\affiliation{Instituto de Telecomunica\c{c}\~oes, Instituto Superior T\'{e}cnico, University of Lisbon, 1049-001 Lisboa, Portugal}
\affiliation{The Blackett Laboratory, Imperial College London, London SW7 2AZ, United Kingdom}
\affiliation{Kavli Institute for Theoretical Sciences, University of Chinese Academy of Sciences, Beijing, China}

\author{M. Matsuo}
\affiliation{Kavli Institute for Theoretical Sciences, University of Chinese Academy of Sciences, Beijing, China}
\affiliation{CAS Center for Excellence in Topological Quantum Computation, University of Chinese Academy of Sciences, Beijing, China}
\affiliation{Advanced Science Research Center, Japan Atomic Energy Agency, Tokai, Japan}
\affiliation{RIKEN Center for Emergent Matter Science (CEMS), Wako, Saitama, Japan}

\author{T. Kato}
\affiliation{Institute for Solid State Physics, The University of Tokyo, Kashiwa, Japan}

\date{\today}

\begin{abstract}
We consider a ferromagnetic particle levitated in air under microwave irradiation and theoretically study the noise in its rigid-body rotation induced by the gyromagnetic effect.
This rotational noise includes useful information on angular momentum transfer from the magnetization to the rigid-body rotation, such as the unit of angular momentum per one spin relaxation process. 
We formulate the rotational noise in terms of the Lindblad equation, which describes the quantum stochastic process, and estimate it in the case of realistic experimental parameters.
We show that a bifurcation phenomenon observed in our setup amplifies the noise and, therefore, can be exploited making an accurate measurement of the rotational noise.
\end{abstract}
\maketitle 

\section{Introduction}
\label{sec:introduction}

Gyromagnetic effects, i.e., conversion between spin and mechanical rotation, has been intensively studied since its first demonstrations by Barnett, Einstein, and de Haas~\cite{Barnett1915,Einstein_deHaas_1915}. While gyromagnetic effects were initially studied in bulk magnetic materials to determine their gyromagnetic ratios~\cite{Scott1962}, they have recently been observed in various condensed matter systems and are now recognized as universal phenomena~\cite{Matsuo-PRL-2011,matsuo2013mechanical,hirohata2018magneto,Takahashi2016,Takahashi2020,Kazerooni2020,Kazerooni2021,Kobayashi2017,kurimune2020highly,tateno2020electrical} and particle physics~\cite{Adamczyk2017,adam2018global,adam2019polarization,acharya2020evidence,adam2021global}. These effects have provided powerful tools for measuring and controlling both mechanical and magnetic properties~\cite{imai2018observation,imai2019angular,Zolfagharkhani-NatNano-2008,Wallis2006,Harii2019,Mori2020,Dornes2019,izumida2022einstein,tateno2021einstein}. The Barnett effect, which converts the mechanical rotation's angular momentum into spin~\cite{Barnett1915}, has been used to generate spin current from different sources~\cite{hirohata2018magneto,Kobayashi2017,Matsuo-PRL-2011,matsuo2013mechanical,kurimune2020highly,tateno2020electrical,Takahashi2016,Takahashi2020,Kazerooni2020,Kazerooni2021} and identify the angular momentum compensation temperature of ferrimagnets~\cite{imai2018observation,imai2019angular}. Conversely, the Einstein-de Haas effect, which generates mechanical torque from spin polarization~\cite{Einstein_deHaas_1915}, has been utilized to measure faint torques caused by single electron spin flips~\cite{Zolfagharkhani-NatNano-2008}, identify the gyromagnetic ratio of nanomagnetic thin films~\cite{Wallis2006}, and analyze the demagnetization process in ferromagnets on sub-picosecond time scales~\cite{Dornes2019}.

\begin{figure}[tbh]
    \centering
    \includegraphics[width=80mm]{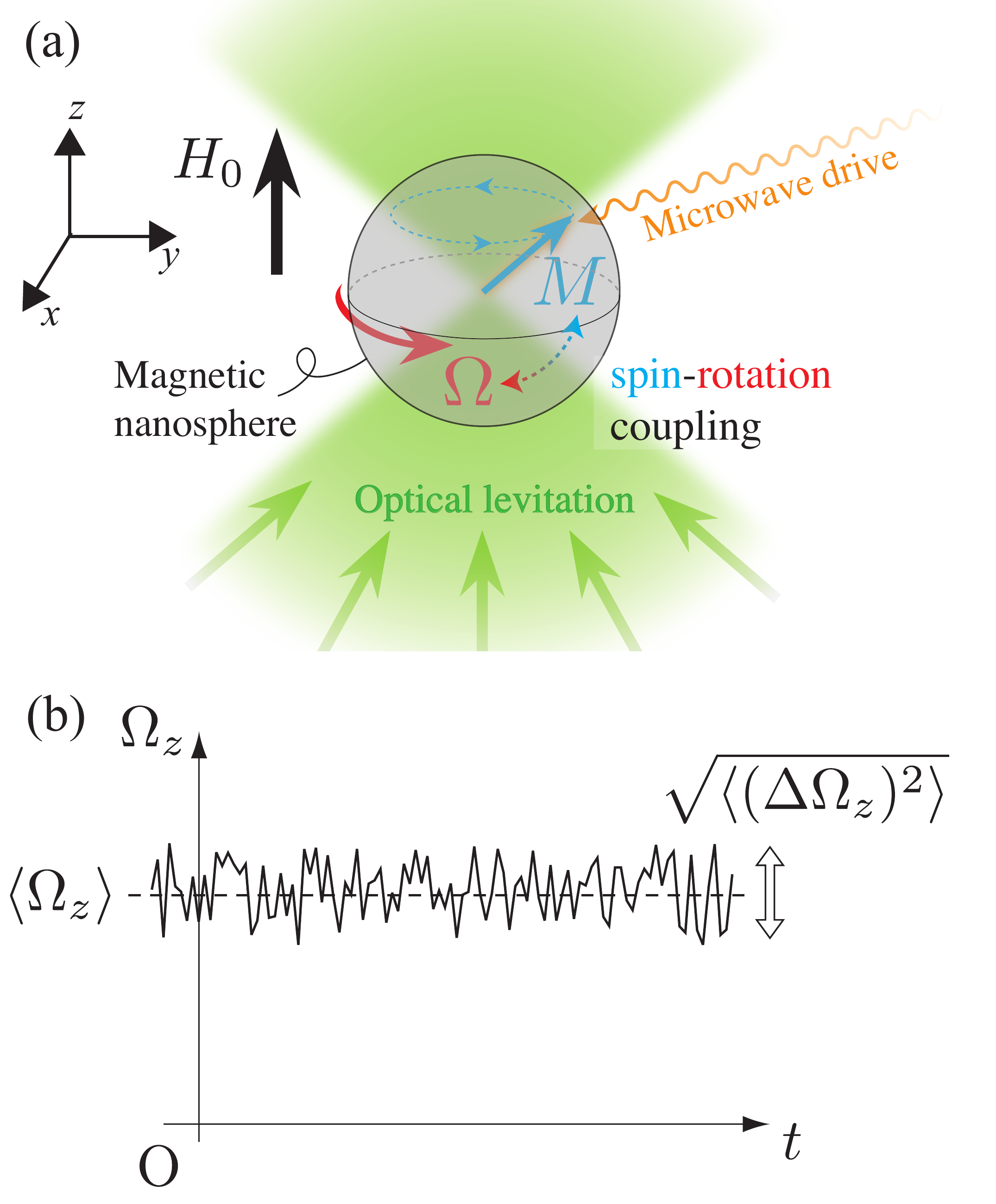}
    \caption{(a) Schematic diagram of magnetic particle levitated by an optical levitation technique.
    Ferromagnetic resonance (FMR) is induced by microwave irradiation under a static magnetic field $H_0$.
    The Gilbert damping of the magnetization of the particle causes the particle to rotate through the spin-rotation coupling.
    (b) Schematic diagram of the noise in the rotational frequency $\Omega_z$.}
    \label{fig:setup}
\end{figure}

Recently, the coexistence of Barnett and EdH effects has been investigated~\cite{tateno2021einstein}. Though there are many ways to investigate gyromagnetic effects, the usual methods target solid systems that are not isolated, for which angular momentum dissipation becomes a problem. Our previous paper~\cite{Sato2023} focused on this issue and proposed to utilize an optical levitation technique~\cite{ashkin1970acceleration,ashkin1976optical,ashkin1986observation,chu1985three,vuletic2000laser,gonzalez2021levitodynamics,li2011millikelvin,gieseler2012subkelvin,jain2016direct,Keshtgar2017,Lyutyy2019} (see Fig.~\ref{fig:setup}(a)). Optical levitation technique enables us not only to make an isolated system but also to reduce friction acting on the particle. Thanks to these advantages, the angular momentum loss can be controlled and the system becomes suitable for surveying gyromagnetic effects. 
We theoretically demonstrated that a swift rotation, which is essential to the coexistence of the above gyromagnetic effects, can be realized by combining optical levitation technique and a ferromagnetic resonance. We showed that the coexisting gyromagnetic effects lead to a bifurcation phenomenon and make possible a precise measurement of the $g$-factor of the spin-rotation coupling. Moreover, our previous work indicated that a combination of an optically levitated system and coexisting of the gyromagnetic effects is a promising platform for researching the gyromagnetic nature of solids. 

In this paper, we focus on fluctuation of the rotation frequency, as shown in Fig.~\ref{fig:setup}(b), instead of steady-state solutions, as discussed in our previous paper~\cite{Sato2023}. 
We show that rotational noise, in particular shot noise, includes useful information on the angular momentum transfer from the magnetization to the rigid-body rotation of the levitated particle, reflecting that magnetization relaxation accompanies a quantum stochastic `kick' exerted on the particle.
We formulate the dynamics of the magnetization in terms of the Lindblad equation, which can describes quantum stochastic processes and the back-action due to quantum measurements in spin systems, while we describe the particle's dynamics by incorporating a `kick' term as well as air resistance terms into the Euler equation.
In the following, we derive and solve these coupled equations around the steady state to compute the fluctuation in the particle's rotation frequency resulting from the `kick'.
We also point out that the bifurcation phenomenon observed in our setup is useful for making an accurate measurement of the rotational noise because it amplifies the noise. 

The remaining part of this paper is organized as follows. In Sec.~\ref{sec:setup}, we summarize the results of our previous paper and explain the bifurcation phenomenon. 
In Sec.~\ref{sec:quantumdescription}, we formulate the fluctuation in the torque of a levitated particle induced by spin relaxation. We introduce the Lindblad equation to describe the stochastic quantum process involved in the spin dynamics. 
In Sec.~\ref{sec:noise}, we calculate the rotational noise by solving the Langevin equation for rigid-body rotation of the particle by using the results of Sec.~\ref{sec:quantumdescription}.
We also estimate the rotational noise and discuss the conditions under which the torque fluctuation can be observed. 
In Sec.~\ref{sec:summary}, we summarize our results.
Two appendices provide the detailed derivations of the equations.

\section{Setup and Steady-State Rotation}
\label{sec:setup}

We consider a spherical ferromagnetic particle of radius $r$ and mass $m_\mathrm{ptc}$ optically levitated in the air (Fig.~\ref{fig:setup}).
The particle is treated as a rigid body with a moment of inertia, $I=2m_\mathrm{ptc} r^2/5$.
We further consider ferromagnetic resonance (FMR) induced by external microwaves~\cite{Kittel1948}.
An external static magnetic field $H_0$ is applied in the $z$ direction and the magnetization of the particle, ${\bm M}$, is initially aligned to the $z$-axis.
In this setup, the angular momentum of the excited spins is transferred to the rigid-body rotation of the particle via the spin-rotation coupling.
This means that the microwave irradiation causes the particle to rotate around the $z$-axis~\footnote{Although the other components of the angular frequency, $\Omega_x$ and $\Omega_y$, should be zero in average, they may fluctuate around zero by thermal noise due to the air resistance.
In order to treat the fluctuations of $\Omega_x$ and $\Omega_y$, we need to perform complex analytic calculation for the LLG and Euler equations in three dimensions~\cite{Sato2023}.
However, this fluctuation includes only the information on thermal fluctuation.
In our work, we focused only on the fluctuation of $\Omega_z$ to study the nonequilibrium noise induced by the Gilbert damping.}
We denote its rotation frequency vector as ${\bm \Omega}=(0,0,\Omega_z)$.

Our previous work~\cite{Sato2023} examined the steady-state rotation, by combining the Landau-Lifshitz-Gilbert (LLG) equation with the equation of motion for rigid-body rotation.
Its results are summarized as follows.
The magnetization is rewritten in terms of the total spin ${\bm S}$ as ${\bm M} = \hbar \gamma \langle {\bm S}\rangle /V$, where we have introduced the gyromagnetic ratio $\gamma=g_e e/2m_e\simeq -1.76\times 10^{11}\,{\rm rad/(s\cdot T)}$.
Note that $e$, $m_e$, and $g_e$ are the charge, mass, and $g$-factor of the electron, and $V=4\pi r^3/3$ is the volume of the particle. 
The Hamiltonian in the rotating frame fixed to the particle is given by~\cite{hehl1990inertial,Frohlich1993,Matsuo2013}
\begin{align}
{\cal H} &= -(\mu_0\gamma{\bm H}+g_{\rm SR} {\bm \Omega}) \cdot \hbar {\bm S},
\label{eq:Hamiltonian} \\
\label{eq:H}
{\bm H}&=\begin{pmatrix}
h\cos(\omega-\Omega_z) t\\
h\sin(\omega-\Omega_z) t\\
H_0
\end{pmatrix},
\end{align}
where $\mu_0\simeq1.257\times10^{-6}\,{\rm N/A^2}$ is the vacuum permeability, ${\bm H}$ is the magnetic field in the rotating frame, $g_{\rm SR}$ is the $g$-factor for the spin-rotation coupling, 
and $h$ ($\ll H_0$) and $\omega$ are the amplitude and frequency of the microwave, respectively. 
The first and second terms of the Hamiltonian describe the Zeeman energy and the spin-rotation coupling, respectively.
The latter term includes the gyromagnetic effects, i.e., the Barnett effect~\cite{Barnett1915} and the Einstein-de Haas effect~\cite{Einstein_deHaas_1915}. 

\begin{figure}
    \centering
    \includegraphics[width=80mm]{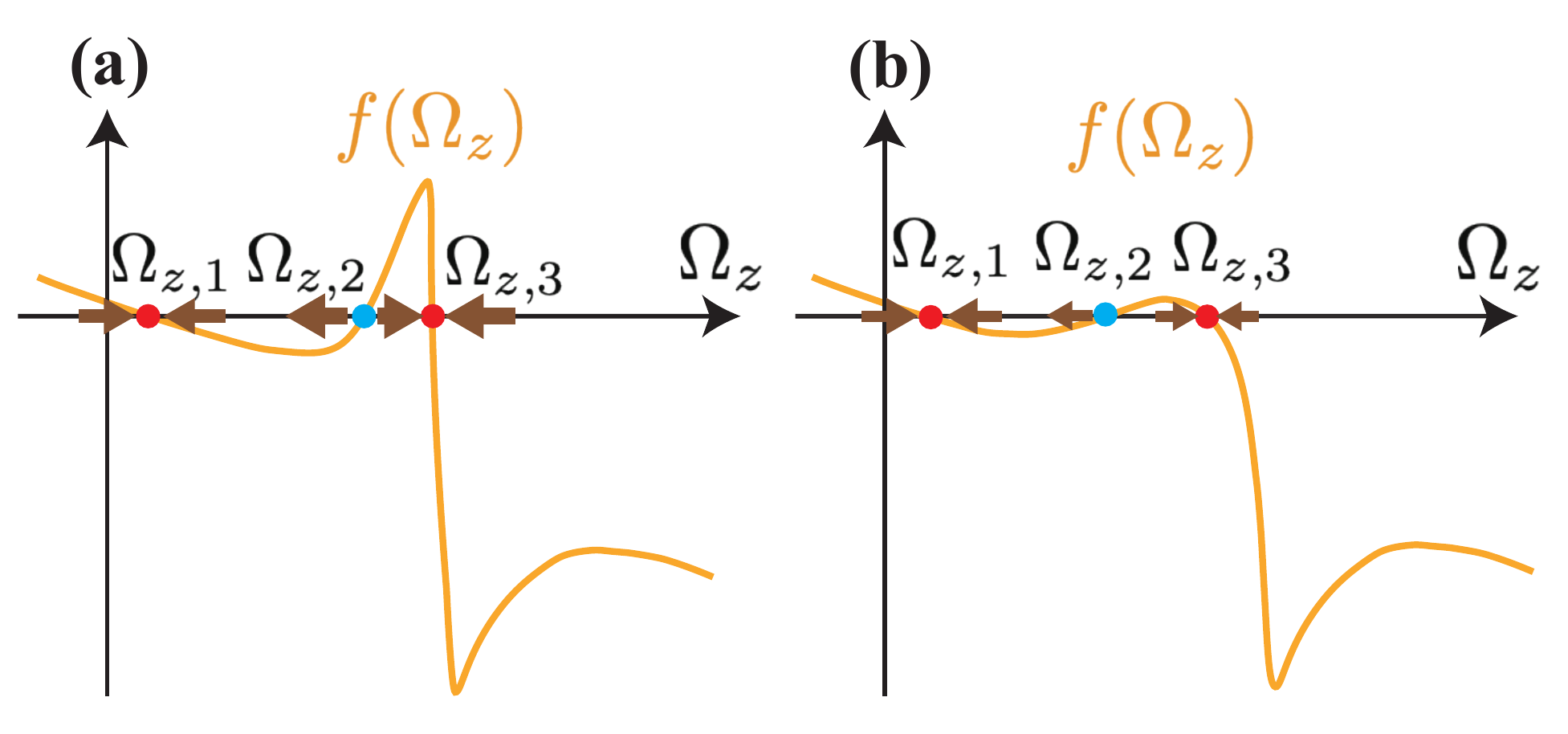}
    \caption{Torque $f(\Omega_z)$ acting on levitated particle as a function of the rotation frequency $\Omega_z$ (a) away from the bifurcation and (b) near the bifurcation.
    The restoring torque toward the stable solutions (indicated by the arrows) becomes weak when the system is near the bifurcation.}
    \label{fig:setup2}
\end{figure}

The Landau–Lifshitz–Gilbert (LLG) equation for the above Hamiltonian is
\begin{align}
\label{eq:LLG}
\dot{{\bm M}} =
{\bm M}\times (\mu_0\gamma {\bm H}+g_{\rm SR}{\bm \Omega})+\frac{\alpha}{M_0}{\bm M}\times
\dot{{\bm M}},
\end{align}
where $M_0=|{\bm M}|$, $\dot{{\bm M}}={\rm d}{\bm M}/{\rm d}t$, and $\alpha$ is the Gilbert damping constant. The steady-state solution of the LLG equation can be obtained by assuming $\dot{M_z}=0$ and $(M_{x},M_y) = M(\cos\left[(\omega -\Omega_z)t+\phi\right],\sin\left[(\omega -\Omega_z)t+\phi\right])$
and using $M_0^2=M^2+M_z^2$. For this steady state, $M_z$ is determined as 
\begin{align}
M_z &= \frac{1}{\sqrt{2}\alpha(\omega-\Omega_z)/M_0} \sqrt{-A + \sqrt{A^2+B}}, \\
A & =\mu_0^2\gamma^2 h^2+(\omega-\omega_0+\Delta g\Omega_z)^2-\alpha^2(\omega-\Omega_z)^2, \\
B & =4\alpha^2(\omega-\Omega_z)^2(\omega-\omega_0+\Delta g\Omega_z)^2,
\end{align}
where $\omega_0= -\mu_0\gamma H_0$ is the resonant frequency and $\Delta g\equiv g_{\rm SR}-1$. 
Note that $g_{\rm SR}$ can deviate from unity due to the spin-orbit coupling in the particle~\cite{Sato2023,Matsuo2013}.
The $z$ component of the angular momentum transferred per unit time from the spin system to the lattice system through Gilbert damping, $\Gamma_{g}$, is given as
\begin{align}
\label{eq:FMR}
\Gamma_{g}&\equiv \frac{\hbar\alpha}{S_0}\left[\langle {\bm S} \rangle \times \langle \dot{\bm S} \rangle \right]_z \nonumber \\
&= -\frac{\alpha V}{M_0\gamma}(\omega-\Omega_z)(M_0^2-M_z^2).
\end{align}
The total torque applied to the particle, $f(\Omega_z)$, is obtained as
\begin{align}
f(\Omega_z)=-\beta\Omega_z + \Gamma_{g},
\end{align}
where the first and second terms describe the angular momentum lost to air resistance and the angular momentum gained from the spin system, respectively.
The air resistance coefficient $\beta$ is given as 
\begin{align}
\beta=-\frac{8 r^4}{3}\sqrt{\frac{\pi m_{\rm air}}{2k_{\rm B}T}}p,
\end{align}
in the molecular flow region~\cite{Epstein1924,Roth-1990-book,Corson2017,Ahn2018}. $k_B$ is the Boltzmann constant, $T$ is the temperature, $m_{\rm air}$ is the average molecular mass, and $p$ is the pressure.
The steady-state solution of $\Omega_z$ is obtained by solving $f(\Omega_z)=0$. 
This equation can be reduced to a cubic equation in $\Omega_z$ and has three solutions at most.
Figure~\ref{fig:setup2}~(a) shows a schematic graph of $f(\Omega_z)$ and the three steady-state solutions.
Two solutions, $\Omega_{z,1}$ and $\Omega_{z,3}$, are stable because an infinitesimal change in $\Omega_{z}$ induces a restoring torque.
On the other hand, the other solution $\Omega_{z,2}$ is unstable because no restoring torque works there.

We observe these bifurcation phenomena~\cite{Sato2023}
when the Gilbert damping constant is large compared to the ratio of the microwave amplitude and the static magnetic field~\footnote{If Eq.~(\ref{eq:bifurcation}) is not satisfied, only one solution for the rotation frequency is physically acceptable for arbitrary pressure, as the other solutions give imaginary values for the magnetization. For details, see Ref.~\cite{Sato2023}.},
\begin{align}
\label{eq:bifurcation}
\frac{h}{H_0} < \frac{\alpha}{2},
\end{align}
and the pressure is taken near 
\begin{align}
p_{\rm 1}=\frac{4|\gamma|\mu_0^2M_0h^2}{r\alpha\omega_0^2}\sqrt{\frac{\pi k_\mathrm{B}T}{2 m_{\rm air}}}.
\label{eq:p1}
\end{align}
Near the bifurcation, the restoring torque is reduced, as schematically shown in Fig.~\ref{fig:setup2}~(b).
This remarkable feature can be used to increase the fluctuation in the rotation frequency around the steady-state solution.

Below, we theoretically consider the fluctuation in the rotation frequency and clarify that it provides important information related to angular momentum transfer from the spin systems to the rigid-body rotation.
In particular, we study the nonequilibrium noise of the torque, which reflects quantum effects such as quantization of the angular momentum transfer and back-action due to the spin relaxation process.
For this purpose, we will formulate the quantum dynamics of the spin in the particle by using the Lindblad equation.

\section{Quantum description of spin dynamics}
\label{sec:quantumdescription}

Although the LLG equation is a basic equation for describing ferromagnetic resonance, it can only treat average values or thermal noise of the magnetization.
To consider the quantum nature of the nonequilibrium fluctuation in the magnetization, we need to introduce a quantum stochastic equation.
For this purpose, we will use the Lindblad equation, which is one of the fundamental equations describing stochastic processes in open quantum systems.
We will formulate spin dynamics in terms of stochastic spin relaxation, which accompanies angular momentum transfer into the rigid-body rotation.

The quantum mechanical interpretation of the LLG equation and its thermal noise has been the subject of several articles~\cite{rebei2003fluctuations,Norambuena2020,yuan2022master,anders2022quantum}. 
Here, we further proceed these discussions by using the Fokker-Planck equation, which enables us to determine the stochastic dynamics of the magnetization under the Gilbert damping. This formulation provides a framework to evaluate the noise induced by the Gilbert damping.

\subsection{Lindblad equation}

The Lindblad equation for the density operator of the total spin $\rho_S$ is given by 
\begin{align}
\label{eq:Lindblad}
\dv{\rho_S}{t} &= -\frac{i}{\hbar}\left[\mathcal{H},\rho_S\right]\nonumber \\
&\hspace{8mm} +\Gamma L \rho_S L^\dagger  -\frac{\Gamma}{2}L^\dagger L \rho_S-\frac{\Gamma}{2} \rho_S L^\dagger L , \\
\mathcal{H} &= -\hbar\mu_0\gamma \left[ H_0' S_z +\frac{h}{2}(S_+ e^{-i\omega' t}+{\rm h.c.} ) \right] ,
\end{align}
where $S_\pm\equiv S_x\pm iS_y$, and we have defined the spin relaxation rate $\Gamma$, the effective magnetic field $H_0'=H_0-g_{\rm SR}\Omega_z$, and the frequency measured in the rotating frame $\omega'=\omega-\Omega_z$. 
The jump process is taken as $L=S_-$ to describe energy relaxation from the spin system into the bath~\footnote{One can consider the other jump operators, $S_+$ and $S_z$, which respectively describes the energy gain from the phonon bath and dephasing due to fluctuation of the Zeeman field.
In our work, we neglect these effects since in the ferromagnetic resonance which drives the magnetization away from thermal equilibrium states, the nonequilibrium noise due to the energy relaxation is expected to be dominant.}.
Hereafter, we will set the $g$-factor of the spin-rotation coupling as $g_{\rm SR}=1$.

Now, let us describe the steady state of the total spin under microwave irradiation.
When the microwave is irradiated for a sufficiently long time, the total spin precesses around the $z$-axis at a constant frequency.
The steady-state dynamics are described by the spin coherent state, defined as
\begin{align}
\ket{\tau e^{i\omega' t}}=\sum_{n=0}^{2S_0} \frac{\sqrt{_{2S_0}C_n} (\tau e^{i\omega' t})^n}{(1+|\tau|^2)^{S_0}} \ket{2S_0-n},
\end{align}
where $\tau$ is a complex number and $S_0$ is the amplitude of the total spin ${\bm S}$.
Using the spin coherent state, the density operator can be expressed in terms of the distribution function $W(\tau,\tau',t)$,
\begin{align}
\label{eq:expansion}
\rho_S=\int d^2\tau W(\tau,\tau^*,t)\ket{\tau e^{i\omega' t}}\bra{\tau e^{i\omega' t}},
\end{align}
where $\tau^*$ is the complex conjugate of $\tau$.
After a lengthy but straightforward calculation~\cite{Takahashi1975}, Eq.~(\ref{eq:Lindblad}) reduces to the Fokker-Planck equation,
\begin{align}
\nonumber
\pdv{W}{t}&=\Bigl\{ \pdv{}{\tau}\left[i(\omega-\omega_0)\tau-i\Omega(\tau^2 -1)-\Gamma(S_0+1)\tau\right]
\\&+{\rm h.c.}+\frac{\Gamma}{2}\pdv[2]{}{\tau}\tau^2+\frac{\Gamma}{2}\pdv[2]{}{{\tau^*}} 
 \tau^{*2}+\Gamma\pdv[2]{}{\tau}{\tau^*}\Bigr\}W,
 \label{eq:FokkerPlankEq}
\end{align}
where $\Omega=-\mu_0\gamma h/2$. Hereafter, we will consider only the leading term of order of $S_0$, assuming that the ferromagnetic particle includes a sufficient number of magnetic atoms.
That is, the last three terms on the r.h.s.~of Eq.~(\ref{eq:FokkerPlankEq}), which are $O(S_0^{-1})$, can be dropped and the distribution function for the steady spin precession state can be written as~\footnote{We chose the sign in front of $e^{i\theta_1/2}$ of Eq.~(\ref{eq:tau1}) plus, because $\langle S_z\rangle\leq 0$ holds in our setup when the magnetic field is applied in the $+z$ direction.}
\begin{align}
\label{eq:stationary}
& W_{st}(\tau,\tau^*,t)=\delta^2(\tau-\tau_1), \\
\label{eq:tau1}
& \tau_1=\frac{\omega-\omega_0+i\Gamma S_0
\pm e^{i\theta/2}\Delta^{1/4}}{2\Omega} , \\
&\Delta =
[(\omega-\omega_0)^2+4\Omega^2-(\Gamma S_0)^2]^2 \nonumber \\
&\hspace{10mm} +4(\Gamma S_0)^2(\omega-\omega_0)^2, \\
& e^{i\theta} = \frac{\left[(\omega-\omega_0)+i\Gamma S_0 \right]^2+4\Omega^2}{\Delta^{1/2}}.
\label{eq:tau1e}
\end{align}
The density operator for this distribution is written as
\begin{align}
\rho_S = \ket{\tau_1 e^{i\omega't}}\bra{\tau_1 e^{i\omega' t}} .
\label{eqtest}
\end{align}

Next, let us discuss how the relaxation rate $\Gamma$ in the Lindblad equation is determined by comparison with the LLG equation.
Multiplying both sides of Eq.~(\ref{eq:Lindblad}) by $S_z$ and $S_\pm$ and taking the trace (see Appendix \ref{appx:Sz,S-} for detailed derivation), we obtain the equations of motion for the expectation values of the spin operators as 
\begin{align}
\label{eq:Sz}
\langle\dot{S}_z\rangle &= -i \Omega \left[ \langle S_+ \rangle e^{-i\omega' t} - \langle S_- \rangle e^{i\omega' t}\right]
\\\nonumber &\hspace{1cm}-\Gamma \langle S_+\rangle \langle S_- \rangle +O(\Gamma S_0),
\\
\label{eq:S+}
\langle\dot{S}_+\rangle&=i\omega_0 \langle S_+\rangle-2i\Omega \langle S_z\rangle e^{i\omega' t}
\\\nonumber &\hspace{1cm}+\Gamma \langle S_z\rangle \langle S_+\rangle+O(\Gamma S_0),
\\
\label{eq:S-}
\langle\dot{S}_-\rangle&=-i\omega_0 \langle S_-\rangle+2i\Omega \langle S_z\rangle e^{-i\omega' t}
\\\nonumber &\hspace{1cm}+\Gamma \langle S_z\rangle \langle S_-\rangle+O(\Gamma S_0).
\end{align}
 On the other hand, the LLG equation~(\ref{eq:LLG}) is rewritten with the expectation values of the spin operators as 
\begin{align}
\langle S_z\rangle &=i\frac{\mu_0\gamma h}{2}\left[\langle S_+ \rangle e^{-i\omega' t}-\langle S_- \rangle e^{i\omega' t}\right] \nonumber \\
&\hspace{1cm}-\frac{\alpha}{S_0}\omega' \langle S_- \rangle \langle S_+ \rangle , \\
\langle \dot{S}_+\rangle&=-i\mu_0\gamma H_0 \langle S_+ \rangle + i\mu_0\gamma h \langle S_z \rangle e^{i\omega' t}\nonumber \\
&\hspace{1cm}+\frac{\alpha}{S_0}\omega'\langle S_z \rangle \langle S_+ \rangle, \\
\langle \dot{S}_-\rangle &=i\mu_0\gamma H_0 \langle S_- \rangle - i \mu_0\gamma h \langle S_z \rangle e^{-i\omega' t}\nonumber \\
&\hspace{1cm}+\frac{\alpha}{S_0}\omega' \langle S_z \rangle \langle S_- \rangle.
\end{align}
By comparison, we can show that the Lindblad equation restores the LLG equation in the leading order of $S_0$ when we take
\begin{align}
\Gamma=\frac{\alpha\omega'}{S_0} .
\label{eq:GammaAlpha}
\end{align}
We can also check that the steady spin precession derived from the LLG equation is consistent with the spin coherent state given in Eq.~(\ref{eqtest}), noting that the expectation value of the spin operator for the coherent state $\ket{\xi}$ is given as
\begin{align}
\mel*{\xi}{S_z}{\xi} &= S_0 \frac{1-|\xi|^2}{1+|\xi|^2}, \\
\mel*{\xi}{S_+}{\xi} &= S_0 \frac{2\xi}{1+|\xi|^2}, \\
\mel*{\xi}{S_-}{\xi} &= S_0 \frac{2\xi^*}{1+|\xi|^2}.
\end{align}
In the following discussion, we will use the Lindblad equation (\ref{eq:Lindblad}) with the condition (\ref{eq:GammaAlpha}) in order to discuss the nonequilibrium nature of the Gilbert damping.

\subsection{Correlation function of a torque}

In order to calculate a correlation function of the torque acting on the particle from the Lindblad equation, we formulate the Poissonian measurement between time $t$ and time $t+dt$.
We define a random variable, $dN(t)$, which takes one if the spin relaxation described by the jump operator $L=S_-$ occurs and takes zero otherwise.
This variable satisfies the following equations:
\begin{align}
\label{eq:dN}
dN(t)dN(t)=dN(t),\qquad 
\langle dN(t)\rangle=dt\langle L^\dag L\rangle.
\end{align}
Since the angular momentum transferred per spin relaxation process is $\hbar$, the torque on the particle, 
$\Gamma_g$, is described as
\begin{align}
\label{eq:Gamma_g}
\Gamma_g(t)=\hbar\frac{dN(t)}{dt}.
\end{align}
By using Eq.~(\ref{eq:dN}), we can show that the expectation value of the torque, $\langle\Gamma_g\rangle$, agrees with Eq.~(\ref{eq:FMR}) for the steady spin precession state.

For $t<t'$, we can write the correlation of the torques as
\begin{align}
\nonumber
&\langle \Gamma_g(t')\Gamma_g(t)\rangle \\
\nonumber &=\frac{\hbar^2}{dt^2}\sum_{dN=\{0,1\}} dN(t')dN(t)  P[dN(t), dN(t')]
\\&=\frac{\hbar^2}{dt^2}\langle dN(t)\rangle\times P[dN(t')=1|dN(t)=1],
\end{align}
where $P[a(t),b(t')]$ is the joint probability of getting the value $a$ at time $t$ and $b$ at time $t'$ and $P[b(t')|a(t)]$ is the conditional probability of getting $b$ at $t'$ under the condition that the result of a measurement at $t$ is $a$. 
The conditional probability $P[dN(t')=1|dN(t)=1]$ is calculated as follows~\cite{risken1996fokker,breuer2002theory,gardiner2004quantum,wiseman2009quantum}.
If the spin relaxation occurs at time $t$, i.e., $dN(t)=1$, the post-measurement state is described in terms of the density operator as
\begin{align}
\rho'_S(t)=\frac{L\rho_S(t) L^\dag}{\langle L^\dag L\rangle}.
\end{align}
By defining a superoperator $\mathcal{L}$ as
\begin{align}
\mathcal{L}\rho_S = -\frac{i}{\hbar}\left[\mathcal{H},\rho_S\right]+\Gamma L \rho_S L^\dagger -\frac{\Gamma}{2} \{ L^\dagger L, \rho_S\},
\end{align}
the density operator at time $t'$ ($>t$) becomes $\rho_S(t')=e^{\mathcal{L}(t'-t)}\rho'_S(t)$. 
Therefore, the conditional probability becomes
\begin{align}
&P[dN(t')=1|dN(t)=1] \nonumber \\
&=dt\Tr\left\{L^\dag L e^{\mathcal{L}(t'-t)}\rho'_S(t)\right\}.
\end{align}
In addition, if $t'=t$, the correlation function becomes
\begin{align}
\expval{\Gamma _ g (t') \Gamma _ g (t)}
&=\frac{\hbar^2}{dt^2} \langle dN(t)^2 \rangle 
= \frac{\hbar^2}{dt}\langle L^\dag L\rangle,
\end{align}
by using Eq.~(\ref{eq:dN}). This equal-time divergence indicates that the torque correlation function includes a delta-function of time as
\begin{align}
\expval{\Gamma _ g (t') \Gamma _ g (t)}& \simeq \hbar^2\langle L^\dag L\rangle \delta(t'-t), 
\end{align}
In summary, the correlation function of the torque is given as
\begin{align}
&\langle \Gamma_g(t')\Gamma_g(t)\rangle \nonumber \\
&=\hbar^2\left[ \Tr\left\{ L^\dag Le^{\mathcal{L}(t'-t)} L\rho_s L^\dag\right\}+\langle L^\dag L\rangle \delta(t'-t)\right].
\end{align}
From the Lindblad equation, we can calculate the first term in the bracket on the r.h.s. as
\begin{align}
\nonumber& \hbar^2 \Tr\left\{ L^\dag Le^{\mathcal{L}(t'-t)} L\rho_S L^\dag\right\}
\\&\hspace{1cm}= \langle \Gamma_g(t) \rangle^2 + F_1(t'-t) + O(S_0^0), 
\label{eq:Tr1}
\end{align}
where 
\begin{align}
F_1(t'-t) &= 16\Gamma^2\frac{S_0^3|\tau_1|^2(1-|\tau_1|^2)}{(1+|\tau_1|^2)^4}e^{-(t'-t)(2\Omega\Im(\tau_1)-\Gamma S_0)} \nonumber \\
& \hspace{5mm}\times \cos\left[\left(-(\omega-\omega_0)+2\Omega\Re(\tau_1)\right)(t'-t)\right] .
\label{eq:Tr2}
\end{align}
See Appendix~\ref{app:spindynamics} for a detailed derivation.
Note that the equation,
\begin{align}
2\Omega\Im(\tau_1)-\Gamma S_0 =-\Gamma \langle S_z\rangle,
\end{align}
holds from Eq.~(\ref{eq:tau1}).
Combining these results, we finally obtain
\begin{align}
&\langle \Gamma_g(t')\Gamma_g(t)\rangle - \langle \Gamma_g(t)\rangle^2 
=F_1(t'-t) + F_2 \delta(t'-t), \nonumber \\
& F_2 = \hbar^2 \Gamma\frac{4S_0^2|\tau_1|^2}{(1+|\tau_1|^2)^2} .
\end{align}
Here, $F_1(t)$ describes spin dynamics from the post-measurement state, while $F_2$ describes `shot noise' that originates from the fact that the angular momentum is transferred in units of $\hbar$.

\section{Fluctuation in rotation frequency}
\label{sec:noise}

\subsection{Formulation}

The dynamics of the spherical particle rotation is determined from the Euler equation with Langevin noise, described as
\begin{align}
\label{eq:Euler}
I\dot{\Omega}_z=\langle\Gamma_{\rm air}(\Omega_z)\rangle+\langle\Gamma_{\rm g}(\Omega_z)\rangle+\xi_{air}(t)+\xi_{\rm g}(t),
\end{align}
where $\langle\Gamma_{\rm air}\rangle\equiv-\beta\Omega_z$ is the torque from air resistance, $\xi_{\rm air}(t)$ and $\xi_{\rm g}(t)$ are the Langevin noises of the torques due to the air resistance and the Gilbert damping, respectively.
These Langevin noises are characterized by
\begin{align}
\langle\xi_{\rm air}(t')\xi_{\rm air}(t)\rangle &= 2D\delta(t'-t), \\
\langle\xi_{\rm g}(t')\xi_{\rm g}(t)\rangle &= \langle\Gamma_{\rm g}(t')\Gamma_{\rm g}(t)\rangle-\langle\Gamma_{\rm g}\rangle^2 \nonumber \\
&= F_1(t'-t) + F_2\delta(t'-t)\label{eq:<xi_g xi_g>},
\end{align}
where $D=\beta k_B T$ is Einstein's coefficient for thermal noise.

Let us focus on the vicinity of the stable solutions and consider the Euler equation linearized with respect to $\Delta \Omega_z = \Omega_z-\langle\Omega_z\rangle$,
\begin{align}
I\dv{}{t}\Delta\Omega_z &= -\epsilon \Delta\Omega_z+\xi_{\rm air}(t)+\xi_{\rm g}(t), \\
\epsilon &\equiv \beta-\left.\pdv{\langle \Gamma_{\rm g}\rangle}{\Omega_z}\right|_{\Delta\Omega_z=0} .
\end{align}
The parameter $\epsilon$, which is positive for stable steady-state solutions, indicates the distance from the bifurcation point.

Applying a Fourier transformation,
\begin{align}
\Delta\Omega_z(t)=\int \frac{d\lambda}{2\pi}\Delta\Omega_{z,\lambda} e^{i\lambda t} ,
\end{align}
the correlation function in the frequency domain can be calculated as
\begin{align}
&\langle\Delta\Omega_{z,\lambda'}\Delta\Omega^*_{z,\lambda}\rangle
=\frac{1}{i\lambda' I + \epsilon  }\frac{1}{-i\lambda I + \epsilon }
\nonumber \\
&\hspace{5mm}\times\int dt \int dt'
\left[\langle \xi_{\rm air}(t') \xi_{\rm air}(t)\rangle + \langle \xi_g(t') \xi_g(t)\rangle\right]
 e^{i\lambda t-i\lambda' t'}.
\end{align}
Thus, we obtain the noise power of the rotation frequency:
\begin{align}
S(f)&\equiv2 \int_{-\infty}^\infty dt e^{-ift}\langle\Delta\Omega_z(t)\Delta\Omega_z(0)\rangle, \nonumber \\
&=\frac{2}{f^2I^2+\epsilon^2} (S_1+S_2+S_{3+}(f)+S_{3-}(f)), \\
S_1 &= 2D = 2\beta k_{\rm B}T, \\
S_2 &= \hbar^2\Gamma \frac{4S_0^2|\tau_1|^2}{(1+|\tau_1|^2)^2}, \\
S_{3\pm}(f) &= -16 \hbar^2 \frac{S_0^4|\tau_1|^2(1-|\tau_1|^2)^2}{(1+|\tau_1|^2)^5} \nonumber \\
&\hspace{-5mm} \times \frac{\Gamma^3}{(f\pm (\omega-\omega_0)-2\Omega\Re(\tau_1))^2 +(\Gamma \langle S_z \rangle)^2}.
\label{eq:S3pm}
\end{align}

In the next section, we will focus on the zero-frequency component of the noise power,
\begin{align}
\label{eq:S0}
S &\equiv \lim_{f\rightarrow 0} S(f) 
= \frac{2}{\epsilon^2} (S_1 + S_2 + S_3), \\
S_3 &= -32 \hbar^2 \Gamma^3  \frac{S_0^4|\tau_1|^2(1-|\tau_1|^2)^2}{(1+|\tau_1|^2)^5} \nonumber \\
&\times \frac{1}{(\omega-\omega_0-2\Omega\Re(\tau_1))^2 +(\Gamma \langle S_z\rangle)^2}.
\label{eq:S3}
\end{align}
We find that as the system approaches the bifurcation point ($\epsilon \rightarrow + 0$), the noise power of the rotation frequency is enhanced due to the prefactor $2/\epsilon^2$.
This means that bringing the system close to the bifurcation point enables more accurate measurements of the rotational fluctuation.
The zero-frequency noise is composed of three contributions: thermal noise $S_1$ induced by the surrounding air, shot noise $S_2$ due to the quantized angular momentum transfer, and back-action noise $S_3$ caused by the quantum measurement of the spin state.
Using Eq.~(\ref{eq:GammaAlpha}), we find that $S_2 \propto \alpha S_0$ and $S_3 \propto \alpha^3 S_0$.
If the Gilbert damping coefficient is sufficiently small ($\alpha \ll 1$), $S_2$ becomes larger than $S_3$.
Let us further focus on the shot noise term $S_2$.
Since the average rotation frequency is
\begin{align}
\langle\Omega_z\rangle=\frac{\langle\Gamma_g\rangle}{\beta}=\frac{\hbar\Gamma}{\beta}\frac{4S_0^2|\tau_1|^2}{(1+|\tau_1|^2)^2} ,
\end{align}
the ratio between the $S_2$ and $\langle\Omega_z\rangle$ becomes
\begin{align}
\label{eq:hbar}
{\cal F} = \frac{S_2}{\langle\Omega_z\rangle}= \frac{\beta}{\epsilon^2}\hbar .
\end{align}
This ratio is an analog of the Fano factor defined for electronic transport and includes information on the unit of the angular momentum transfer, $\hbar$.
In order to measure this shot-noise contribution, we need to sufficiently suppress the thermal noise from the air, i.e., to realize a situation in which $S_1 \lesssim S_2$.

\subsection{Numerical estimate}

Let first us discuss the conditions under which to measure the contribution of the shot noise to the fluctuation in the rotation frequency.
Consider the ratio between the thermal noise $S_1$ and the shot noise $S_2$:
\begin{align}
\frac{S_2}{S_1}
=\frac{\hbar\langle\Omega_z\rangle}{2 k_B T}.
\end{align}
To increase this ratio, a high rotation frequency or low temperature is required.
In our estimate, we set the temperature as $T=3 \,{\rm K}$ and the magnetic field as $H_0= 2.6\times 10^6\, {\rm A/m}$ to obtain a high-speed rotation.
Moreover, we set the Gilbert damping constant as $\alpha = 6.7\times 10^{-5}$ and the magnitude of the magnetization as $M_0=1.557\times 10^5\,{\rm A/m}$~\cite{Kajiwara2010}.
The amplitude of the microwave and the air pressure are $h=10\,{\rm A/m}$ and $p=p_1\simeq 3.4\times 10^{-5} \, {\rm Pa}$, respectively [see Eq.~(\ref{eq:p1})].

\begin{figure}
    \centering
    \includegraphics[width=80mm]{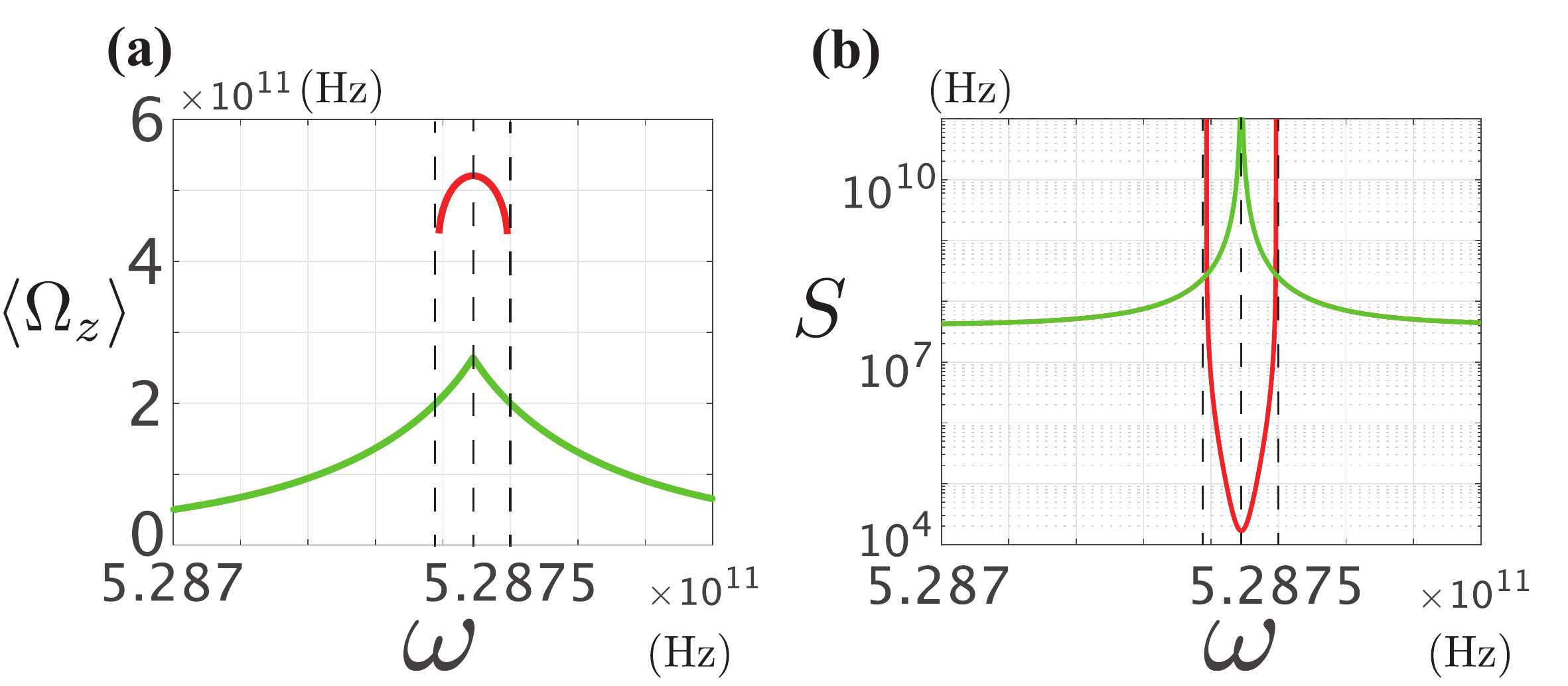}
    \caption{(a) Rotation frequency and (b) zero-frequency noise power of the 
    stable steady-state solutions as a function of microwave frequency. The parameters are set as $h=10\,{\rm A/m}$ and $p=p_1$. 
    Note that (b) is a semi-log plot. 
    The vertical dashed lines indicate the bifurcation points, at which $\epsilon$ becomes zero.}
    \label{fig:result}
\end{figure}

Figures~\ref{fig:result}~(a) and (b) show the average and zero-frequency noise power of the rotation frequency.
The red and green lines correspond to the two stable solutions with high and low rotation frequencies.
In Fig.~\ref{fig:result}~(a), the endpoints of the red curve and the central cusp of the green curve correspond to the bifurcation point ($\epsilon \rightarrow +0$).
Correspondingly, the zero-frequency noise $S$ diverges there as shown in Fig.~\ref{fig:result}~(b), because of the factor $2/\epsilon^2$ in Eq.~(\ref{eq:S0}). 
This result indicates that the noise of the rotation frequency is amplified near the bifurcation.

\begin{figure}
    \centering
    \includegraphics[width=50mm]{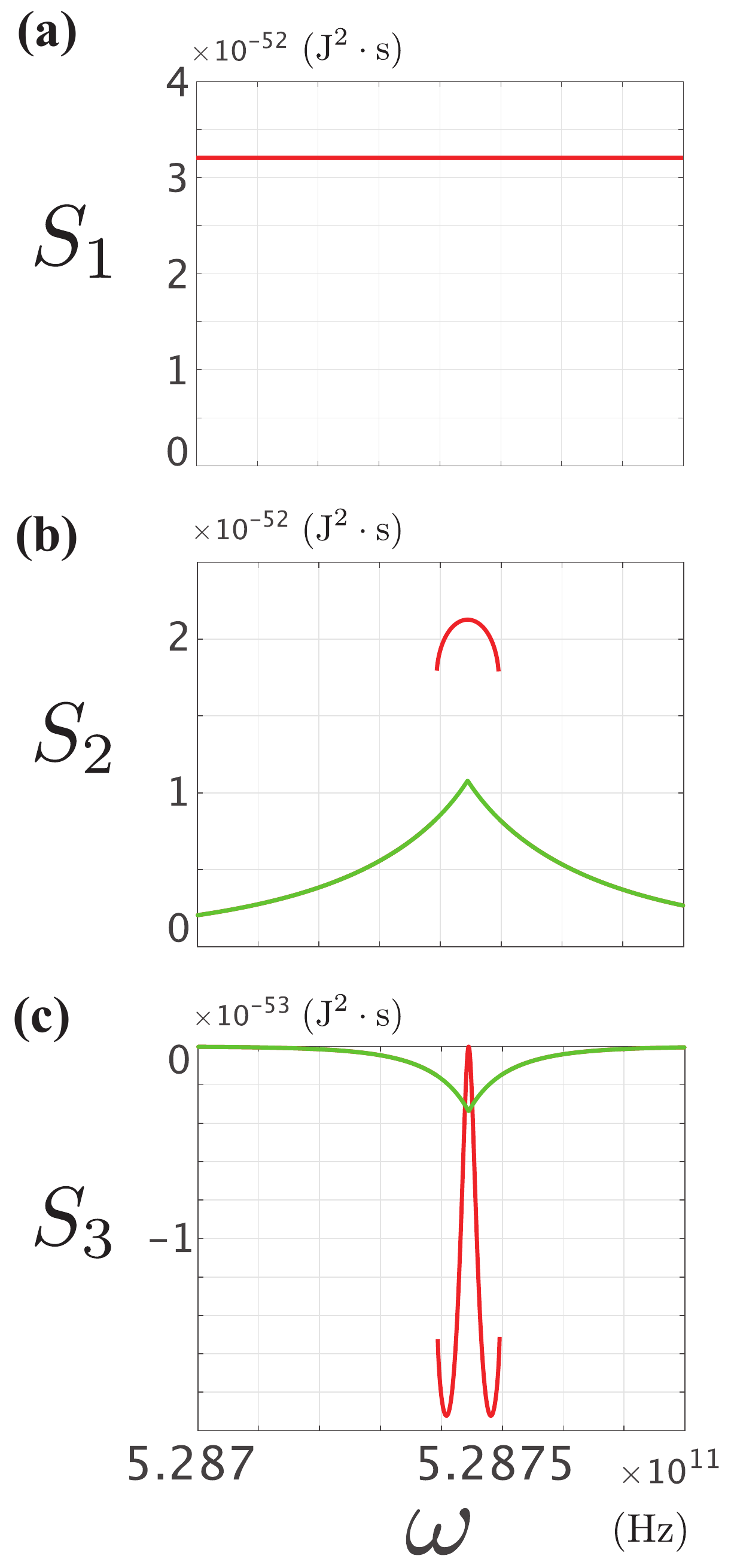}
    \caption{(a) Thermal noise $S_1$ due to air resistance, (b) shot noise $S_2$ due to quantized angular momentum transfer, and (c) back-action noise $S_3$ induced by quantum measurement of the spin state as functions of microwave frequency.
    Note that the factor $2/\epsilon^2$ has been eliminated (see Eq.~(\ref{eq:S0})).}
    \label{fig:result2}
\end{figure}

Figures~\ref{fig:result2}~(a), (b), and (c) show the three contributions ($S_1$, $S_2$, and $S_3$) to the noise power as a function of the microwave frequency $\omega$.
Note that there appears no divergent behavior for these noises as the factor $2/\epsilon^2$ has been eliminated [see Eq.~(\ref{eq:S0})].
We find that the thermal noise $S_1$ is independent of the microwave frequency while $S_2$ and $S_3$ significantly depend on it.
This feature will be useful for subtracting the thermal noise from the measured noise power.
For the present parameters, the shot noise term $S_2$ is a little smaller than $S_1$, but is large enough to be measured.
On the other hand, the back-action term $S_3$ is negative and an order of magnitude smaller than $S_1$ and $S_2$.

Finally, we should comment on the magnitude of the standard deviation of the rotation frequency. 
When the measurement interval is given as $\tau_{\rm m}$, the measurement data of the rotation frequency are $\Omega_z' \simeq\frac{1}{\tau_{\rm m}}\int_{t-\tau_{\rm m}/2}^{t+\tau_{\rm m}/2} dt'\,\Omega_z(t')$. 
The zero-frequency noise $S$ can be related to the standard deviation of $\Omega_z'$ as
\begin{align}
S \simeq 2\int_{-\tau_{\rm m}/2}^{\tau_{\rm m}/2} dt'  \langle \Delta \Omega_z(t')\Delta\Omega_z(0) \rangle  \simeq 2\tau_{\rm m} \langle (\Delta \Omega_z')^2 \rangle,
\end{align}
when $\tau_{\rm m}^{-1}$ is sufficiently small compared with the characteristic frequency in the noise power spectrum (see Eq.~(\ref{eq:S3pm})).
Considering that $S\approx 10^{10}\,{\rm Hz}$ and $\langle\Omega_z'\rangle\approx 10^{11}\,{\rm Hz}$ in a typical setup, when we set the measurement interval as $\tau_{\rm m}=1\,\mu{\rm s}$, the rotation frequency fluctuates as 
\begin{align}
\langle \Omega_z'\rangle \pm \sqrt{\langle (\Delta \Omega_z')^2\rangle} \approx 10^{11}\pm 10^8\,{\rm Hz} .
\end{align}
This estimate indicates that the rotational fluctuation is sufficiently detectable~\cite{liu2020robust,van2020optically}.

In summary, our estimate indicates that the shot noise contribution $S_2$ is sufficiently large compared with the thermal noise $S_1$ due to the air resistance and can be measured at low temperatures for a rapidly rotating particle. 
In our estimate, the back-action term $S_3$ becomes an order of magnitude smaller than $S_1$ and $S_2$.
The observation of the back-action noise will be more difficult than that of the shot noise, because it requires nearly two digits of precision for the noise power measurement, which is almost the best accuracy at present~\cite{van2020optically}.
We expect that a highly accurate noise measurement will clarify the properties of the back-action noise.

\section{Summary}
\label{sec:summary}

We considered the ferromagnetic resonance in the levitated magnetic particle and calculated the fluctuation in the rotation frequency $\Omega_z$. The fluctuation comes from air-resistance noise and Gilbert-damping noise. The latter cannot be treated with the classical LLG equation. To resolve this difficulty, we derived the Lindblad equation which reproduces the LLG equation in the limit of $S_0\rightarrow\infty$, where $S_0$ is the total amplitude of all the spins in the ferromagnetic particle. We reduced the Lindblad equation to the Fokker-Planck equation by using the method of the spin coherent state, and obtained the steady-state solutions. Next, we interpreted the Gilbert damping as a Poissonian process and obtained the noise of the Gilbert damping. Finally, we derived the rotation frequency noise from the Langevin equation. 

We showed that the rotation frequency noise is enhanced near the bifurcation. 
In this sense, the bifurcation is useful for making an accurate measurement of the spin transfer noise. 
The rotational frequency noise is composed of three contributions: thermal noise induced by surrounding air, shot noise, and back-action noise.
The latter two noises are induced by the fluctuation in the torque on the particle due to spin relaxation and include detailed information on the angular momentum transferred from the magnetization.
In particular, the shot noise contribution offers information regarding a fundamental unit of angular momentum transfer, $\hbar$.
We showed that the shot noise can be observed by lowering the temperature or increasing the static magnetic field.
Our work provides a powerful method to obtain detailed information on angular momentum transfer from magnetization to rigid-body rotation, which has not been measured so far.
The back-action noise, which is induced by quantum measurement on the spin, was shown to be small in the present setup. A detailed discussion on this type of noise will be given elsewhere.
In this work, we only studied spin relaxation by considering the jump operator $L=S_-$ in Eq.~(\ref{eq:Lindblad}).
In order to describe thermal fluctuations and derhaphing effect, we need to introduce other jump operators, $S_+$ and $S_z$. 
Such extended analysis will also be given elsewhere.

\begin{acknowledgments}
We thank JSPS KAKENHI for Grants (No.~JP20K03831, No.~JP21K03414, No.~21H01800, No.~21H04565, No.~23KJ0702, and No.~23H01839). 
M.M.~is partially supported by the Priority Program of the Chinese Academy of Sciences, Grant No.~XDB28000000.
D.O.~is supported by JSPS Overseas Research Fellowship, by the Institution of Engineering and Technology (IET), and by Funda\c{c}\~ao para a Ci\^encia e a Tecnologia and Instituto de Telecomunica\c{c}\~oes under project UIDB/50008/2020.
T.S.~was supported by the Japan Society for the Promotion of Science through the Program for Leading Graduate Schools (MERIT).
\end{acknowledgments}

\appendix

\section{Spin dynamics}
\label{app:spindynamics}

In this appendix, we derive the expression for $\Tr\left\{ L^\dag Le^{\mathcal{L}(t'-t)} L\rho_S L^\dag\right\}$ given in Eqs.~(\ref{eq:Tr1}) and (\ref{eq:Tr2}).
We first rewrite the Fokker Plank equation (\ref{eq:FokkerPlankEq}) as $\partial W/\partial t = \mathcal{G} W$, where 
\begin{align}
\mathcal{G} &= \pdv{}{\tau'}\Bigl[ i\Omega (1+\tau_1^2)+(i(\omega-\omega_0)-2i\Omega\tau_1-\Gamma S_0)\tau') \Bigr] \nonumber \\
&\hspace{5mm} + {\rm h.c.} 
\end{align}
Note that the diffusion terms (the last three terms in the brackets of Eq.~(\ref{eq:FokkerPlankEq})) are neglected and we have assumed that $\tau$ is sufficiently close to the steady-state $\tau_1$.
If the initial distribution function is given as $W_0(\tau',\tau'^*)=\delta^2(\tau'-\tau)$, its time evolution is obtained as
\begin{align}
& W(\tau',\tau'^*,t) \equiv e^{\mathcal{G}t}W_0 \nonumber \\
&=\delta^2\left(\tau'-\tau_1- \left(\tau-\tau_1\right)e^{-\left[i(\omega-\omega_0)-2i\Omega\tau_1-\Gamma S_0\right]t}\right) .
\label{eq:Wdynamics}
\end{align}
Using Eq.~(\ref{eq:expansion}) and the equation~\cite{Takahashi1975}
\begin{widetext}
\begin{align}
&S_-\ket{\tau e^{i(\omega-\Omega_z) t}}\bra{\tau e^{i(\omega-\Omega_z) t}}S_+ \nonumber 
\\&\simeq\left(\frac{4S_0^2|\tau|^2 
}{(1+|\tau|^2)^2}+\frac{2S_0\tau}{1+|\tau|^2}\pdv{}{\tau}+\frac{2S_0\tau^*}{1+|\tau|^2}\pdv{}{\tau^*}+\pdv[2]{}{\tau^*}{\tau}\right) \ketbra{\tau e^{i(\omega-\Omega_z) t}}{\tau e^{i(\omega-\Omega_z) t}} ,
\end{align}
we can derive the following equation:
\begin{align}
S_-\rho_S(t)S_+ \nonumber&\simeq \int d^2\tau\left\{\left[\frac{4S_0^2|\tau|^2 
}{(1+|\tau|^2)^2}-\pdv{}{\tau}\frac{2S_0\tau}{1+|\tau|^2}-\pdv{}{\tau^*}\frac{2S_0\tau^*}{1+|\tau|^2}+\pdv[2]{}{\tau^*}{\tau}\right]W(\tau,\tau^*,t)\right\}
\\&\hspace{6cm}\times\ket{\tau e^{i(\omega-\Omega_z) t}}\bra{\tau e^{i(\omega-\Omega_z) t}} .
\end{align}
Using $L=S_-$, $\Tr\left\{ L^\dag Le^{\mathcal{L}(t'-t)} L\rho_S L^\dag\right\}$ can be calculated as
\begin{align}
\label{eq:calculation}
\nonumber
\Tr\left\{ L^\dag Le^{\mathcal{L}(t'-t)} L\rho_S L^\dag\right\} &\simeq \Gamma^2\Tr\left\{ \int d^2\tau\left[\left(\frac{4S_0^2|\tau|^2 
}{(1+|\tau|^2)^2}-\pdv{}{\tau}\frac{2S_0\tau}{1+|\tau|^2}-\pdv{}{\tau^*}\frac{2S_0\tau^*}{1+|\tau|^2}+\pdv[2]{}{\tau^*}{\tau}\right)W_{st}(\tau,\tau^*,t)\right]\right.
\\ \nonumber &\left.\hspace{2cm}\times S_+ S_-\int d^2\tau'\delta^2(\tau'-\tau)e^{\mathcal{L}(t'-t)}\ket{\tau' e^{i\omega' t}}\bra{\tau' e^{i\omega' t}}  \right\}
\\\nonumber&=\Gamma^2 \int d^2\tau\left[\left(\frac{4S_0^2|\tau|^2 
}{(1+|\tau|^2)^2}-\pdv{}{\tau}\frac{2S_0\tau}{1+|\tau|^2}-\pdv{}{\tau^*}\frac{2S_0\tau^*}{1+|\tau|^2}+\pdv[2]{}{\tau^*}{\tau}\right)W_{st}(\tau,\tau^*,t)\right]
\\\nonumber&\hspace{2cm}\times \int d^2\tau'\bra{\tau' e^{i\omega' (t+(t'-t))}} S_+ S_-\ket{\tau' e^{i\omega' (t+(t'-t))}} e^{\mathcal{G}(t'-t)}\delta^2(\tau'-\tau) 
\\\nonumber&\simeq\Gamma^2 \int d^2\tau\left[\left(\frac{4S_0^2|\tau|^2 
}{(1+|\tau|^2)^2}-\pdv{}{\tau}\frac{2S_0\tau}{1+|\tau|^2}-\pdv{}{\tau^*}\frac{2S_0\tau^*}{1+|\tau|^2}+\pdv[2]{}{\tau^*}{\tau}\right)\delta^2(\tau-\tau_1)\right]
\\&\hspace{2cm}\times \int d^2\tau'\frac{4S_0^2|\tau'|^2 
}{(1+|\tau'|^2)^2} e^{\mathcal{G}(t'-t)}\delta^2(\tau'-\tau) ,
\end{align}
\end{widetext}
up to the leading order of $S_0$. Finally, we can use Eq.~(\ref{eq:Wdynamics}) to derive Eqs.~(\ref{eq:Tr1}) and (\ref{eq:Tr2}) in the main text.

\section{Derivation of Eq.~(\ref{eq:Sz})-(\ref{eq:S-})}\label{appx:Sz,S-}

In this appendix, we supplement the derivation of Eq.~(\ref{eq:Sz}). Eqs.~(\ref{eq:S+}) and (\ref{eq:S-}) can be derived in the same manner.
By multiplying Eq.~(\ref{eq:Lindblad}) with the spin operator $S_z$ and taking the trace, we obtain
\begin{align}
\langle\dot{S}_z\rangle &= -i \Omega \left[ \langle S_+ \rangle e^{-i\omega' t} - \langle S_- \rangle e^{i\omega' t}\right]
-\Gamma \langle S_+ S_- \rangle.
\label{eq:appB1}
\end{align}
Here, we have used the commutation relation of the spin operator, $[S_i,S_j]=i\epsilon_{ijk}S_k$, with the Levi-Civita symbol $\epsilon_{ijk}$.
Using Eq.~(\ref{eqtest}), the last term of Eq.~(\ref{eq:appB1}) is calculated as
\begin{align}
\langle S_+ S_- \rangle = \langle S_+\rangle\langle S_-\rangle + \frac{2S_0}{(1+|\tau_1|^2)^2}.
\end{align}
The first term on the r.h.s is $O(S_0^2)$, while the second term is $O(S_0)$. Thus, $\langle S_+ S_- \rangle$ of Eq.~(\ref{eq:appB1}) can be replaced with $\langle S_+\rangle\langle S_-\rangle$ in the leading order of $S_0$, to arrive at Eq.~(\ref{eq:Sz}).

\bibliography{ref}

\end{document}